\documentclass[aps,twocolumn,pra,showpacs]{revtex4-1}
\usepackage{graphicx,epstopdf}
\usepackage{amsmath}
\usepackage{amssymb}
\usepackage{color}
\usepackage{float}
\usepackage{epsfig}
\usepackage{epstopdf}
\usepackage{subfigure}
\usepackage{enumitem}

\usepackage[colorlinks=true, citecolor=blue, urlcolor=blue ]{hyperref}
\input{epsf}
\begin{document}

\title{Response to defects in multipartite and bipartite entanglement\\ of isotropic quantum spin networks}

\author{Sudipto Singha Roy$^{1}$, Himadri Shekhar Dhar$^{1,2}$, Debraj Rakshit$^{1,3}$, Aditi Sen(De)$^{1}$, and Ujjwal Sen$^{1}$}

\affiliation{$^1$Harish-Chandra Research Institute, HBNI, Chhatnag Road, Jhunsi, Allahabad 211 019, India, \\
$^2$Institute for Theoretical Physics, Vienna University of Technology (TU Wien), Wiedner Hauptstraße 8-10/136,
A-1040  Vienna, Austria, \\
$^3$Institute of Physics, Polish Academy of Sciences, Aleja Lotnik{\'o}w 32/46, PL-02668 Warsaw, Poland}


\begin{abstract}
Quantum networks are an integral component in performing efficient computation and communication tasks that are not accessible using classical systems. A key aspect in designing an effective and scalable quantum network 
is generating entanglement between its nodes, which is robust against defects in the network. We consider an isotropic quantum network of spin-1/2 particles with a finite fraction of defects, where the corresponding wave function of the network is rotationally invariant under the action of local unitaries.
By using quantum information-theoretic concepts like strong-subadditivity of von Neumann entropy and 
approximate quantum telecloning, 
we prove analytically  that in the presence of defects, caused by loss of a finite fraction of spins, the network,  comprised of a fixed numbers of lattice sites, sustains genuine multisite entanglement, and at the same time may exhibit finite moderate-range bipartite entanglement, in contrast to the network with no defects.
%
\end{abstract}
\maketitle

\section{Introduction}
A critical aspect in the study of quantum networks \cite{kimble,qi} is the distribution of entanglement \cite{horo} between the nodes \cite{adv,adv1,array,mbqc,amico-08,opt-lattice,morton,nv,sq,llyod,horo, long-range}.
For implementation of quantum protocols such as information transmission \cite{array}, long-range quantum teleportation in spin chains \cite{long-range}, and measurement-based quantum computation \cite{mbqc},
engineered generation and modulation of entanglement between the spins on the lattice is a necessary prerequisite. In several hybrid quantum networks designed using superconducting or optomechanical cavities
\cite{hybrid}, entanglement is the key resource enabling the fidelity and speed of information transfer \cite{speed} within the network. 
Hence, 
robustness of entanglement in the presence of  defects is an important 
requirement in the design of scalable computation and quantum-information theoretic (QIT) models.

In this paper, we investigate the consequences of \textit{defects} on the entanglement properties of a quantum spin network with a fixed number of lattice sites.
Such defects may dismantle quantum correlations including entanglement in a system, and hence 
can adversely affect its computational and communication abilities \cite{decoh}.
We consider 
a quantum spin network consisting of spin-1/2 particles on a \textit{bipartite} lattice, with an isotropic topology and of arbitrary dimensions. The interaction between the spins is such that the wave function of the spin network consists of superpositions of all short-range dimer coverings on the lattice \cite{him-prl}. The proposed spin network is  closely related to the dimer models considered  for  quantum computing  model comprised of sequential measurements of individual spins~\cite{mbqc,resource,review} and  fault-tolerant topological quantum computation~\cite{kit2003,nayak2008,wilz,sondhi, topo, res-topo,fault}. In recent years,  attempts have also been made to study the above dimer models using tensor network formalism~\cite{cirac2}, which unveils many important physics related to it. Moreover,  proposal for engineering such isotropic models on spin lattices have also been reported~\cite{dimer_realization1,dimer_realization2,ol}.

We show, using QIT properties such as quantum telecloning \cite{telecl} and the 
strong-subadditivity of von Neumann entropy ($\mathcal{S}_\textrm{VN}$) \cite{von} that even in the presence of a finite fraction of defects, the spin network sustains a considerable amount of genuine multisite entanglement (GME). Moreover, in contrast to defectless lattices, the presence of defects may also generate small but finite bipartite entanglement (BE) between two moderately distant sites. We also discuss the relevance of such spin networks in contemporary quantum computation.

\begin{figure}[h]
 \includegraphics[angle=0,width=4.0cm]{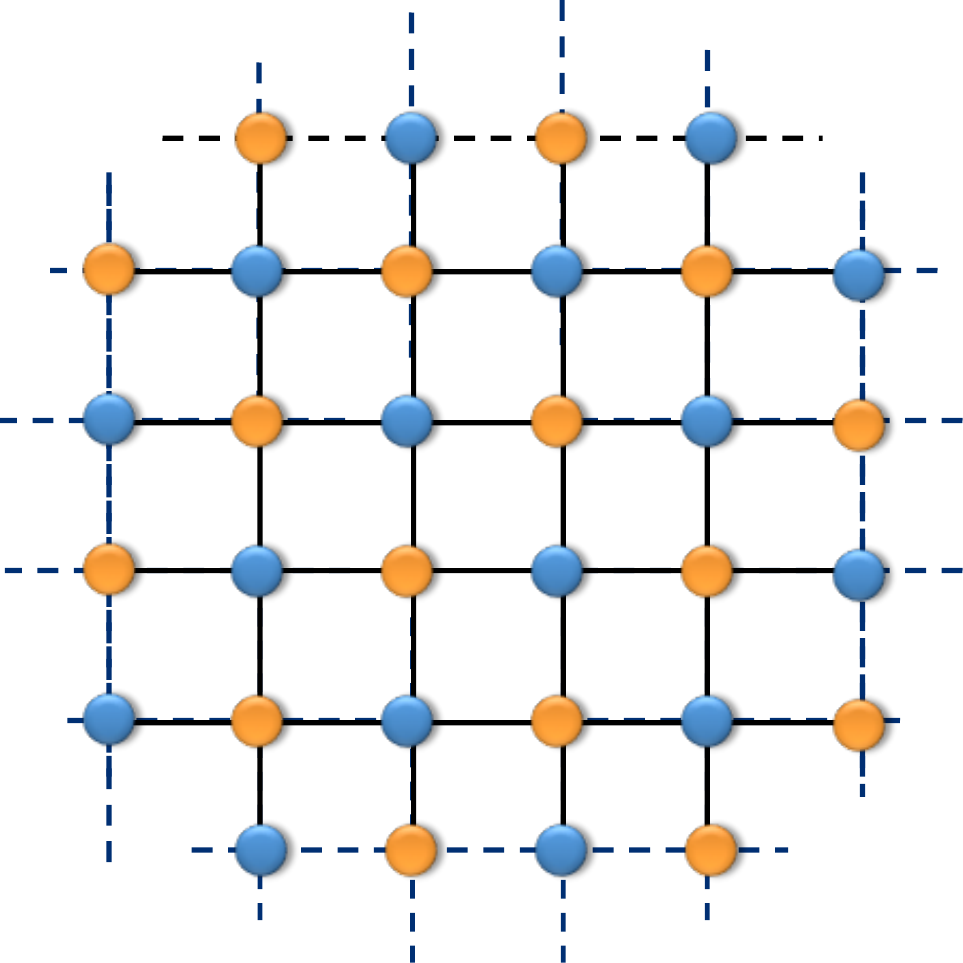} \\
\caption{(Color online). A two-dimensional square bipartite lattice, with sublattice $\mathcal{A}$ (blue-circles) and $\mathcal{B}$ (yellow-circles). The square bipartite lattice can be generalized to a three-dimensional cubic bipartite lattice.}
\label{square}
\end{figure}

The paper is arranged as follows. In Sec.~\ref{isotropic_network}, we introduce the isotropic quantum network of spin-1/2 particles with a finite fraction of defects which we have considered in this work. In Sec.~\ref{red_den}, using the rotational invariance property of the proposed spin network under the action of local unitaries, we derive the mathematical expressions of the few-sites reduced density matrices. Thereafter, in Sec.~\ref{bipartite entanglement}, using the form of the few-site density matrices, we study the bipartite entanglement properties of the doped isotropic network. In Sec.~\ref{multiartite entanglement}, we move one step further and using approximate quantum telecloning and subadditivity of von Neumann entropy, study the multiparty entanglement properties of the above doped spin network. We conclude in Sec.~\ref{discussion}.

\begin{figure}[t]
\subfigure[]{\includegraphics[scale=.15]{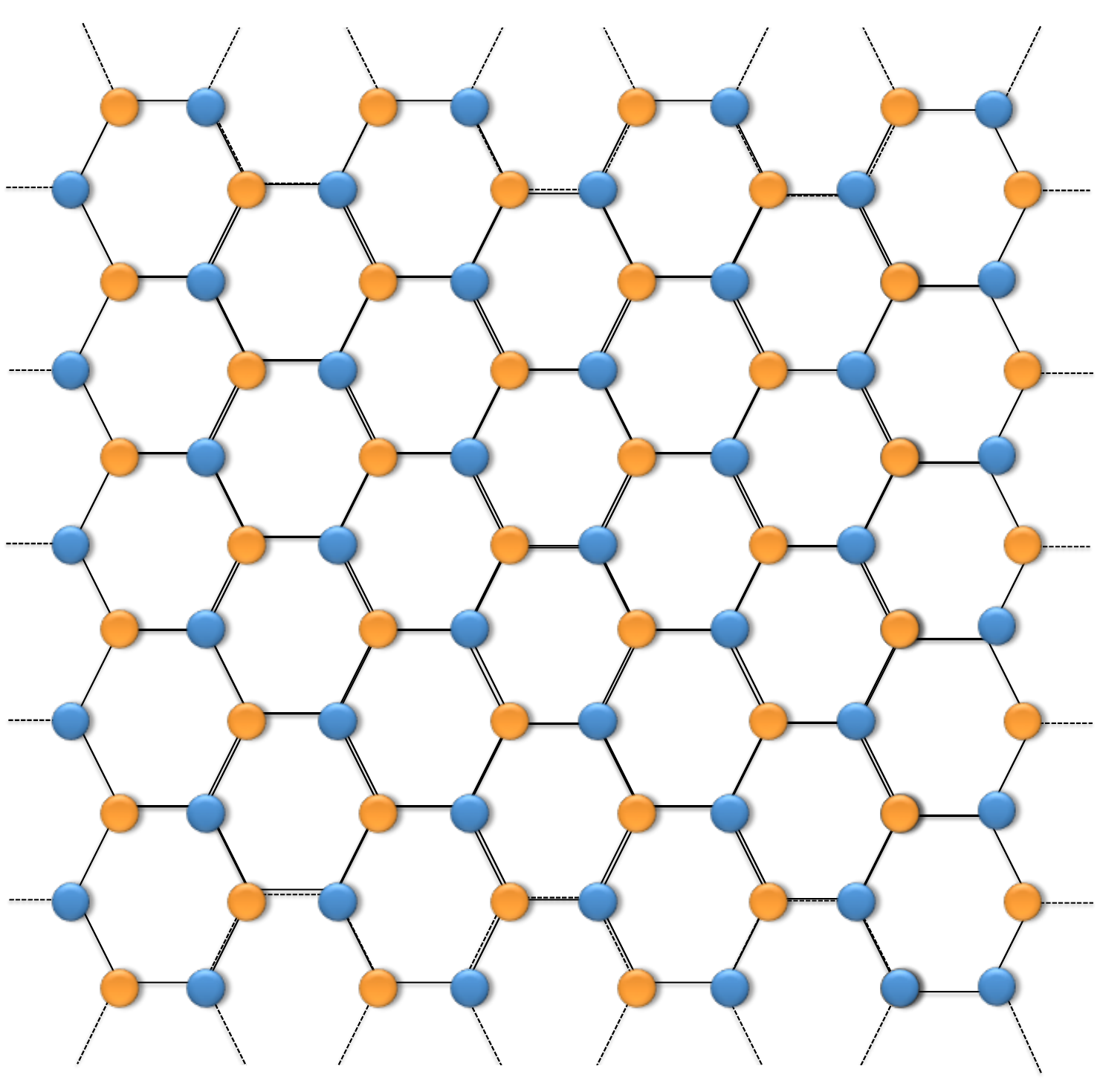} }~~
\subfigure[]{\includegraphics[scale=.15]{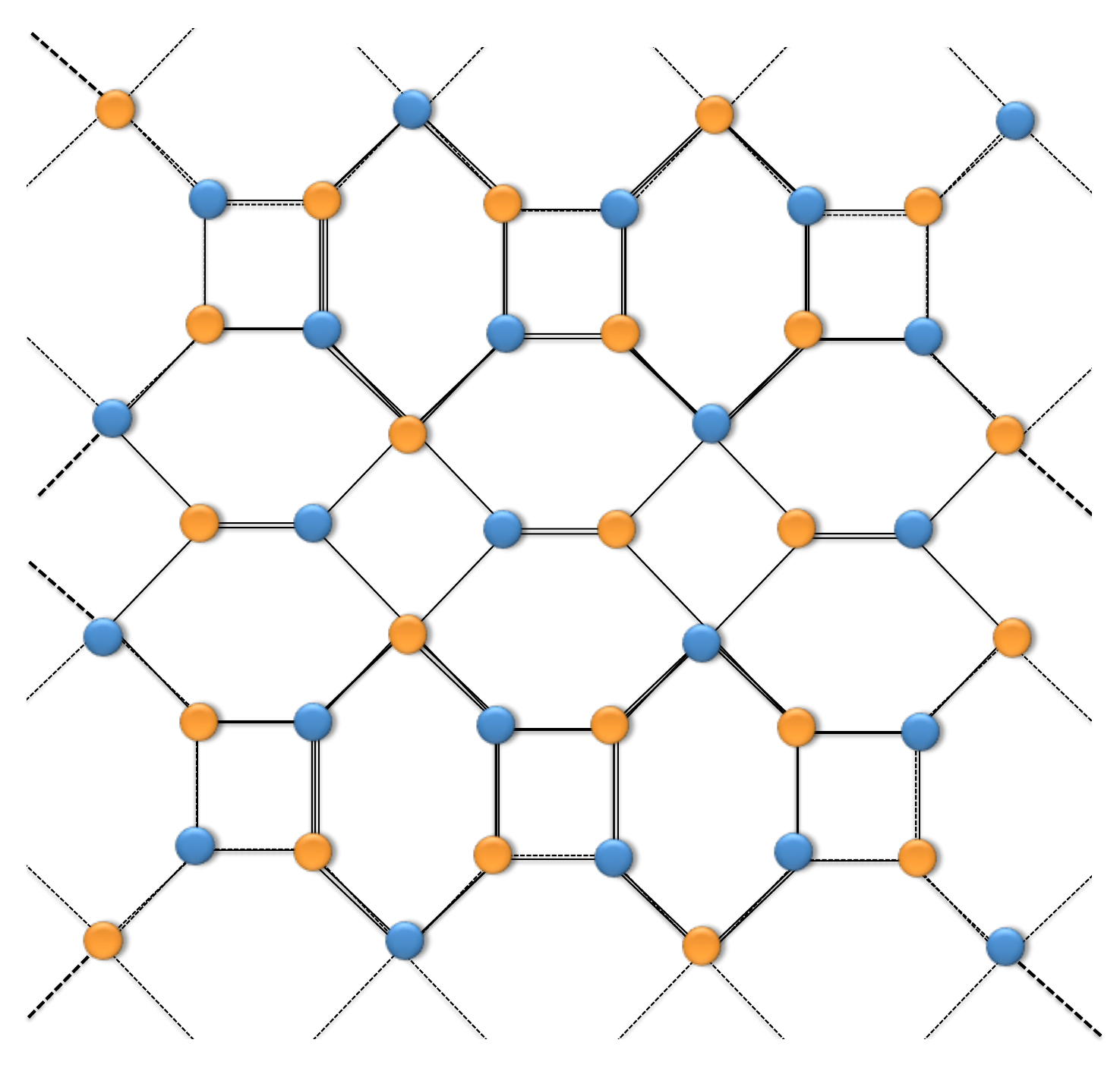}}
\caption{(Color online). A two-dimensional bipartite (a) honeycomb lattice and (b) a dimer lattice of regular polygons, with sublattice $\mathcal{A}$ (blue-circles) and $\mathcal{B}$ (yellow-circles). }
\label{honey}
\end{figure} 


\section{\label{isotropic_network} Isotropic  spin network}
We begin by considering a quantum network consisting of $\mathcal{N}$ (even) lattice sites shared between the two sublattices of a bipatite lattice, 
say $\mathcal{A}$ and $\mathcal{B}$. The lattice is ``bipartite'' in the sense that each site on sublattice \(\mathcal{A}\) is surrounded only 
by sites of sublattice \(\mathcal{B}\), which are 
\(\mathcal{R}\) in number, and vice-versa. 
An example of a bipartite graph or lattice is the two-dimensional square lattice (see Fig.~\ref{square}).
 At this stage, we consider a  situation in which each 
site of the above bipartite lattice is occupied by a spin-1/2 particle. The network is isotropic, which implies that the lattice appears the same from the perspective of any lattice site.
These bipartite graphs are not restricted to two-dimensional square lattices, but can be generalized to the three-dimensional cubic lattice or other two-dimensional graphs such as the honeycomb lattice or other periodic abstract patterns using regular polygons, as shown in Fig.~\ref{honey}.

Moreover, the quantum state of the network is invariant, upto a global phase, under identical local unitaries, i.e., the spin state is \emph{rotationally invariant}.
Such a spin network is potentially equivalent to the spin liquid phases of certain antiferromagnetic strongly-correlated systems \cite{net, ours, aditi-prl}, 
and the state consists of equal-weight superposition of all possible nearest-neighbor (NN) dimer coverings. 
A dimer between any two NN spin-1/2 particles on a bipartite lattice is given by 
$|\psi\rangle_{ij}$ =  $\frac{1}{\sqrt{2}}(|\uparrow_i \downarrow_j \rangle - |\downarrow_i \uparrow_j\rangle)$, 
where $i \in \mathcal{A}$ and $j \in \mathcal{B}$. The (unnormalized) state of the $\mathcal{N}$-spin quantum network can then be defined as
$|\Psi\rangle_\mathcal{N}$ = $\sum_k \{\prod_{i\in\mathcal{A},j\in\mathcal{B}}|\psi\rangle_{ij}\}_k$,
where 
$\{\boldsymbol{\cdot}\}_k$ is the $k^{th}$ (defect-free) dimer covering, which is a product of   
$\mathcal{N}/2$ dimers across the bipartite lattice.\

 Since $|\psi\rangle_{ij}$ is rotationally invariant under operations of the form 
$\mathcal{U}$ $\otimes$ $\mathcal{U}$, where $\mathcal{U}$  is a unitary acting on a single spin-1/2 particle, 
the entire state, $|\Psi\rangle_\mathcal{N}$, is rotationally invariant under the local operation, 
$\mathcal{U}^{\otimes \mathcal{N}}$. In presence of defects,
the description of the state can be mapped to a tensor-product over three-level or  qutrit Hilbert spaces~\cite{map}, such that each node of the network is represented by $\{|\nu_0\rangle,|\nu_1\rangle\}$, where $|\nu_0\rangle$ denotes a node with no spin particle (a defect).

The occupied node is $|\nu_1\rangle$, representing the spin-1/2 particle with 
the two-level basis $\{|\uparrow\rangle, |\downarrow\rangle\}$.
Hence, the overall state of the network can be expressed by local three-level bases, 
$\{|\nu_0\rangle, |\nu_1\rangle|\uparrow\rangle, |\nu_1\rangle|\downarrow\rangle\} \equiv \{|\zeta_0\rangle, |{\zeta}^{\prime}_1\rangle, |{\zeta}^{\prime}_2\rangle\}$. 
In this new basis, a dimer between two NN spins is  $|\psi\rangle_{ij}$ =  $\frac{1}{\sqrt{2}}(|{\zeta}^{\prime}_1\rangle_i |{\zeta}^{\prime}_2\rangle_j - |{\zeta}^{\prime}_2\rangle_i |{\zeta}^{\prime}_1\rangle_j)$, and a defect at node $l$ is written as $|{\zeta}_0\rangle_l$. 
We note that the (un)primed elements of the basis represent the spin (un)occupied sites, corresponding to ($|\nu_0\rangle$)$|\nu_1\rangle$.


\begin{figure}[h]
\subfigure[]{\includegraphics[scale=.17]{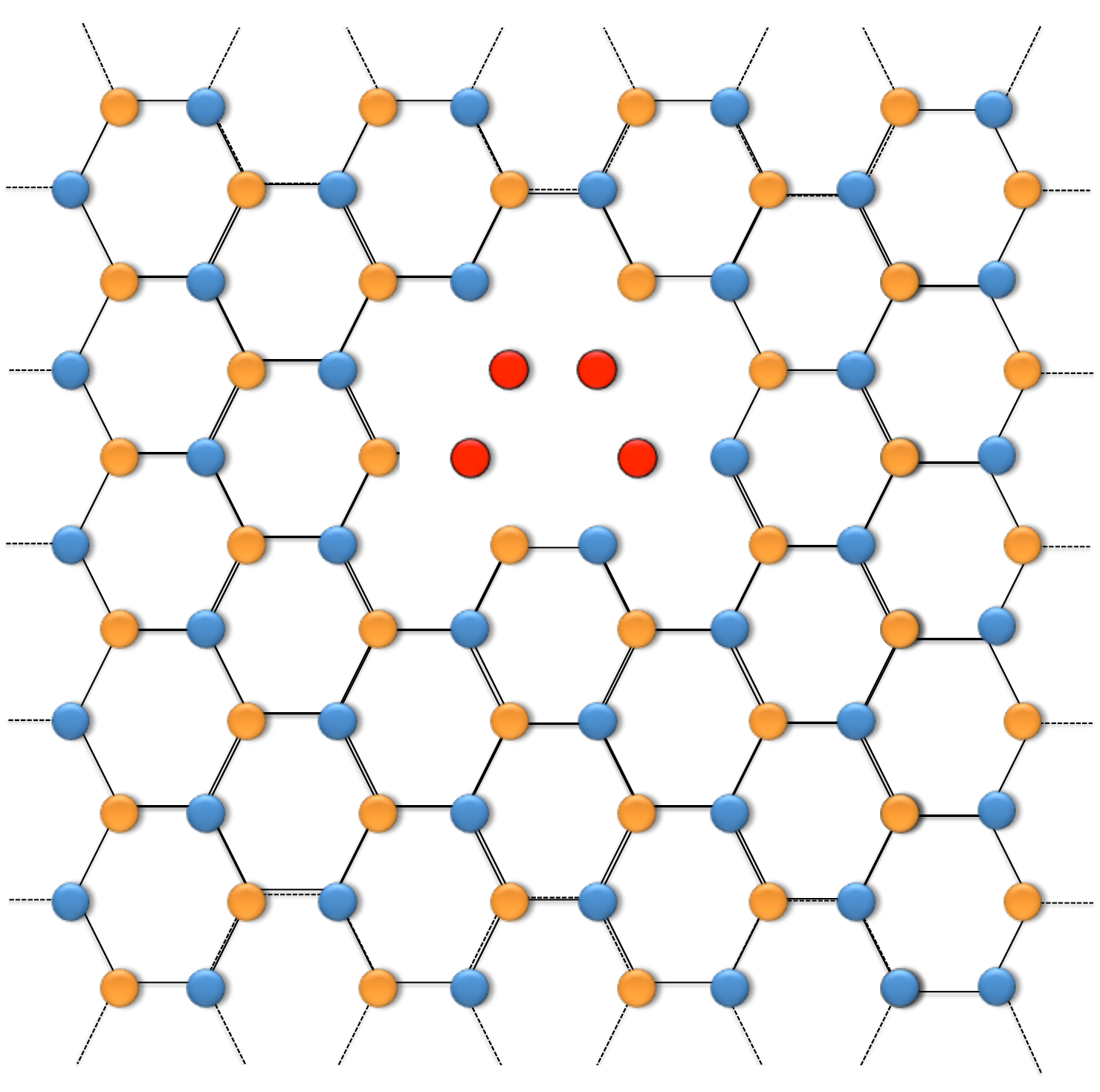} }~~
\subfigure[]{\includegraphics[scale=.17]{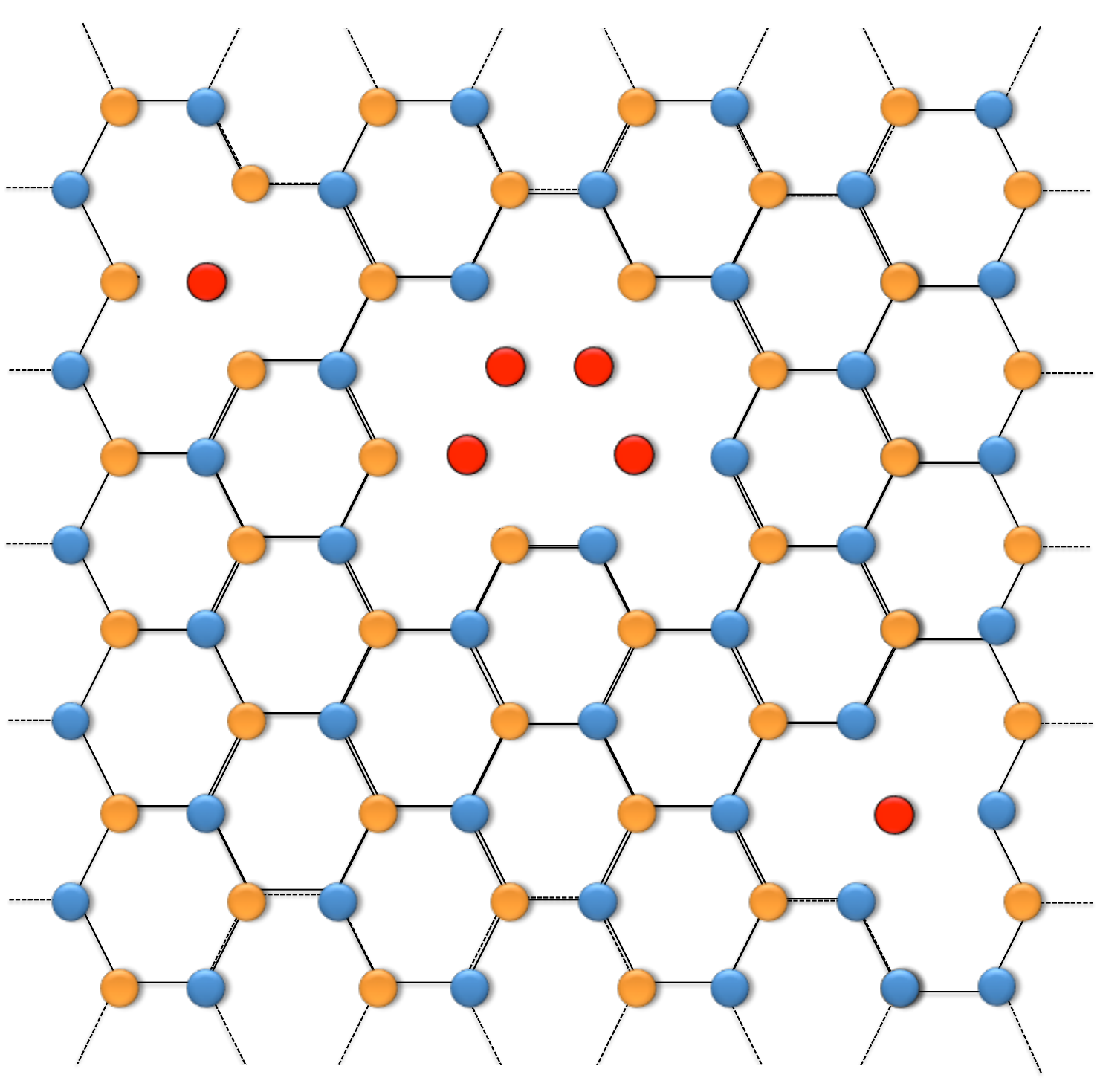}}
\caption{(Color online). Two examples of possible coverings in a two-dimensional bipartite honeycomb lattice, with 
defects (red-circles) in the network. The defects may be (a) clustered or (b) distributed in the lattice, as long as the bipartite state remains intact. The nodes in
sublattice $\mathcal{A}$ (blue-circles) and $\mathcal{B}$ (yellow-circles), are occupied by spin-1/2 particles.}
\label{holes}
\end{figure}

The defects are even in number and are distributed equally between the two sublattices (see Fig.~\ref{holes}). 
An important property of the spin network is its invariance under certain local unitaries. 
For a fixed number of defects at \emph{fixed sites}, 
the spin-occupied sites are always considered to form a complete dimer covering. This is supported by the fact that in 
strongly-correlated systems such coverings are known to be energetically favorable as compared to states where dimer pairs break to form free spins \cite{net}. 
Hence, for an $\mathcal{N}$-site qutrit network, with a fixed number \(\mathcal{P}\) of defects at arbitrary sites of the network, 
the overall state can be written as 
$\bar{\Psi}_{\mathcal{N}}^{\mathcal{P}}$ = $\sum_l |\Psi\rangle^l_{\mathcal{N} - \mathcal{P}} \otimes |\{\phi_l\}\rangle\),
where $|\{\phi_l\}\rangle$  = $|\zeta_0\rangle_{l_1}|\zeta_0\rangle_{l_2}\cdots|\zeta_0\rangle_{l_\mathcal{P}}$ and \(l=\{l_1, l_2, \ldots, l_{\mathcal{P}}\}\). 
The sum in the state is over all \(l_1,l_2, \ldots, l_{\mathcal{P}}\), for a fixed \(\mathcal{P}\).  Here, \(|\Psi\rangle^l_{\mathcal{N} - \mathcal{P}}\) is the state \(|\Psi\rangle_{\mathcal{N} - \mathcal{P}}\) for the \(\mathcal{N}\)-site lattice with the 
sites \(l = \{l_1, l_2, \ldots, l_{\mathcal{P}}\}\) removed.
$\mathcal{D}$ = $\mathcal{P}/\mathcal{N}$ gives us the defect density of the network.
In general, in the presence of defects (where certain spins are missing) or under doping, the ensuing defect-dimer covering not only forms an overcomplete basis, but it is also computationally intractable to enummerate the number of such coverings. This is due to the fact that counting of monomer-dimer pairs on a lattice is NP-complete\cite{jerrum}.\

 In the forthcoming section we will use the invariance property of the doped spin=1/2 network to derive the mathematical expressions of the few-site density matrices.

\section{Derivation of the few-site reduced density matrices}
\label{red_den}
The state of the quantum network with finite defects, $\bar{\Psi}_{\mathcal{N}}^{\mathcal{P}}$, 
is invariant under $\tilde{\mathcal{U}}^{\otimes \mathcal{N}}$ \(\equiv\) $(1 \oplus \mathcal{U})^{\otimes \mathcal{N}}$ = 
$
{\left( \begin{array}{cc}
1 & 0 \\
0 & \mathcal{U}
\end{array}\right)}^{\otimes \mathcal{N}}$, where $\mathcal{U}$ is an arbitrary $2\times2$ unitary operator acting on the spin space 
$\{|\zeta^\prime_1\rangle,|\zeta^\prime_2\rangle\}$.
%
The invariance of $\bar{\Psi}_{\mathcal{N}}^{\mathcal{P}}$  is significant in analyzing the entanglement properties of the quantum network. 
It can be shown that the reduced state, $\rho^{(x)}$,  for any $x$ nodes (lattice sites), obtained by tracing over all but $x$ 
 nodes (say, $\bar{x} = \mathcal{N} - x$)  
from the state $\bar{\Psi}_{\mathcal{N}}^{\mathcal{P}}$,
is also invariant under the action of $\tilde{\mathcal{U}}^{\otimes x}$.
To elaborate a little, 
$
\rho^{(x)} = \textrm{Tr}_{\bar{x}}[|\bar{\Psi}\rangle\langle\bar{\Psi}|] = \sum_{\bar{x}} |\langle\phi_{\bar{x}}|\bar{\Psi}\rangle|^2, 
$
where $\{|\phi_{\bar{x}}\rangle\}$ forms a complete basis over the system of $\bar{x}$ sites, and 
where we have suppressed the sub- and super-scripts of \(|\bar{\Psi}\rangle_{\mathcal{N}}^{\mathcal{P}}\). Now, $\rho^{(x)}$ = $\sum_{\bar{x}}|\langle\phi_{\bar{x}}|\tilde{\mathcal{U}}^{\otimes \mathcal{N}}|\bar{\Psi}\rangle|^2$, due to the invariance property of $|\bar{\Psi}\rangle\langle\bar{\Psi}|$.  Therefore, we have
\begin{eqnarray}
\rho^{(x)} &=& \sum_{\bar{x}}|\langle\phi_{\bar{x}}|\tilde{\mathcal{U}}^{\otimes \bar{x}}\tilde{\mathcal{U}}^{\otimes x}|\bar{\Psi}\rangle|^2 
= \tilde{\mathcal{U}}^{\otimes x}(\sum_{\bar{x}}|\langle\tilde{\phi}_{\bar{x}}|\bar{\Psi}\rangle|^2) \tilde{\mathcal{U}}^{\otimes x \dag}, \nonumber
\end{eqnarray}
where $|\tilde{\phi}_{\bar{x}}\rangle = \tilde{\mathcal{U}}^{\otimes \bar{x} \dag} |{\phi}_{\bar{x}}\rangle $ forms another basis of the system of $\bar{x}$ nodes.
The invariance of  $\rho^{(x)}$ allows us to obtain the expressions for the reduced states of single and two nodes.

For instance, consider the single-node reduced state, $\rho^{(1)}$, obtained by tracing out all but one site from 
\(\bar{\Psi}_{\mathcal{N}}^{\mathcal{P}}\). 
Using the considerations in the preceding paragraph,  we  have $\tilde{\mathcal{U}}\rho^{(1)}\tilde{\mathcal{U}}^\dag$ = $\rho^{(1)}$ \(\forall \tilde{\mathcal{U}}\).
Let $\rho^{(1)}$ = $(\rho^{(1)}_{ij})$ and $\tilde{\mathcal{U}}$ = $(u_{ij})$, where $i,j \in$  $\{1,2,3\}$. 
Now from definition of $\tilde{\mathcal{U}}$, $u_{1j}$ = $u_{j1}$ = $\delta_{1j}~\forall j$, and the unitarity of $\mathcal{U}$ in the spin space 
demands $u_{kl}u^*_{k'l}$ = $\delta_{kk'}$, for $k,k',l \in$ = $\{2,3\}$.
Now $(\tilde{\mathcal{U}}\rho^{(1)}\tilde{\mathcal{U}}^\dag)_{ij}$ = 
$\sum_{j', i' =1}^3 u_{ij'}\rho^{(1)}_{j'i'}u^*_{ji'}$, 
and the invariance is satisfied if
\begin{eqnarray}
\rho^{(1)}_{ij} &=& \sum_{j', i' =2,3} u_{ij'}\rho^{(1)}_{j'i'}u^*_{ji'} + \sum_{i'=1}^3\delta_{i1} \rho^{(1)}_{1i'}u^*_{ji'} \nonumber\\
&+& \sum_{j'=2,3} u_{ij'}\rho_{j'1}^{(1)} \delta_{j1}.
\end{eqnarray}
Hence, 
for all single-node reduced states, the invariance holds if and only if $\rho^{(1)}_{ij}$  is diagonal, 
i.e., $\rho^{(1)}_{ij}$ = $ {\tilde{p}}_i~\delta_{ij}$, with $\sum_i \tilde{p}_{i}$ = 1 and \(\tilde{p}_i \ge 0\), and moreover $\tilde{p}_2$ = $\tilde{p}_3$.
Therefore, the single site reduced density matrix at site $a$ can be written as  
\begin{eqnarray}
\rho^{(1)}_{a} = \mathrm{diag}\{p_1,p_2/2,p_2/2\} = p_1|\zeta_0\rangle\langle\zeta_0|_{a} + p_2 \frac{\mathbb{I}_2}{2},
\label{single}
\end{eqnarray}
with \(p_1,p_2 \ge 0\) and \(p_1+p_2=1\), 
where $\mathbb{I}_2$ = $(0 \oplus {I}_2)$, with ${I}_2$ being the identity matrix in the spin space $\{|\zeta^\prime_1\rangle_{a},|\zeta^\prime_2\rangle_{a}\}$.
Note that the single-site density matrix in the absence of defects is just \(\mathbb{I}_2/2\).
Similarly, one can obtain the analytical expression for the reduced two-node (two-qutrit) states invariant under the local unitary operation, 
$\tilde{\mathcal{U}} \otimes \tilde{\mathcal{U}}$, $\forall~ \tilde{\mathcal{U}}$. 
For an arbitrary pair of nodes $a$ and $b$, where $a\in\mathcal{A}$ and $b\in\mathcal{B}$, the so-deduced 
reduced density matrix is given by
\begin{eqnarray}
\rho^{(2)}_{ab} = p^\prime_1|\zeta_0\zeta_0\rangle\langle\zeta_0\zeta_0|_{ab} + p^\prime_2~ \mathbb{I}^\prime_4/4 + p^\prime_3~ \mathcal{W}(q),
\label{two}
\end{eqnarray}
where $p^\prime_i \geq$ 0 $\forall~i$, and $\sum_ i p^\prime_i$ = 1. The diagonal matrix $\mathbb{I}^\prime_4$, is given by
\begin{eqnarray}
\mathbb{I}^\prime_4 &=& \sum_i |\zeta_0\zeta_{i}^{\prime}\rangle\langle\zeta_0\zeta_{i}^{\prime }|_{ab}+|\zeta_{i}^{\prime}\zeta_0\rangle\langle\zeta_{i}^{\prime}\zeta_0|_{ab}\nonumber\\
&=& (|\zeta_0\rangle \langle \zeta_0|)_a \otimes (\mathbb{I}_2)_b + (\mathbb{I}_2)_a \otimes (|\zeta_0\rangle \langle \zeta_0|)_b.
\end{eqnarray}
The term $\mathcal{W}(q)$, projected to  $\mathbb{C}^2 \otimes \mathbb{C}^2$, where $\mathbb{C}^2$ is spanned by $\{|\zeta^\prime_1\rangle,|\zeta^\prime_2\rangle\}$, 
is the Werner state \cite{wer}, so that in \(\mathbb{C}^3 \otimes \mathbb{C}^3\) it is given by
\begin{eqnarray}
\mathcal{W}(q) = q |\psi\rangle\langle\psi|_{ab} + (1-q)~ \mathbb{I}_4/4,
\end{eqnarray}
for $-1/3 \leq q \leq 1$, with 
$|\psi\rangle_{ab}$ =  $\frac{1}{\sqrt{2}}(|{\zeta}^{\prime}_1 {\zeta}^{\prime}_2\rangle - |{\zeta}^{\prime}_2 {\zeta}^{\prime}_1\rangle)_{ab}$ 
being the spin dimer and $\mathbb{I}_4$ = \(\mathbb{I}_2 \otimes \mathbb{I}_2\) =  $\sum_{i,j=1}^2 |\zeta_{i}^{\prime}\zeta_{j}^{\prime}\rangle\langle\zeta_{i}^{\prime}\zeta_{j}^{\prime}|_{ab}$ 
being the identity matrix in the two-party spin space.
%

Hence, Eqs.~(\ref{single}) and (\ref{two}) give us the analytical expressions for the single- and two-node (qutrit and two-qutrit) 
reduced density matrices of the spin network with finite defects represented by the state $\bar{\Psi}_\mathcal{N}^\mathcal{P}$. 
We will show that QIT tools can help to obtain the bounds on $p^{'}_i$'s. We observe that $\rho^{(1)}_{a}$ and $\rho^{(2)}_{ab}$ are dependent on the defect density, $\mathcal{D}$ = $\mathcal{P}/\mathcal{N}$. 
For instance, for $\mathcal{D}$ = 0, which corresponds to a spin network with no defects, $p_2$ = 1, $p^\prime_3$ = 1. 
For large networks with a small number of defects, $\mathcal{D} \ll 1$, $p_1 \ll p_2$, and $p^\prime_3 \gg p^\prime_1,p^\prime_2$. 
However, exact values of $p_i$'s, $p^\prime_i$'s, and $q$ are difficult to compute and are intractable, even for simple 2D lattices. 
This is due to the fact that even the enumeration of arrangements in a defect-dimer covering is known to be \textit{NP-complete} \cite{jerrum}. 
This makes the estimation of entanglement properties of these spin networks a formidable task. We show that by using QIT concepts 
one can unearth substantial details about the bipartite and multipartite entanglements of 
these quantum spin networks.

\section{Bipartite entanglement properties}
\label{bipartite entanglement}
In this section, we  investigate the BE properties of the state, given by Eq.~(\ref{two}), between two arbitrary sites ($a\in\mathcal{A}$ and $b\in\mathcal{B}$) of the quantum spin network. 
The condition of positive partial transposition (PPT) \cite{ppt} is given by $[p^\prime_3(1-3q)]/4 \geq$ 0. Therefore, $\rho^{(2)}_{ab}$ has a 
non-positive-partial-transpose (NPT), and hence is entangled, iff $q > 1/3$, which, incidentally,
 is the same criterion as for the Werner state to be entangled. To obtain a 
stricter
criterion for the BE of the spin network with defects, and in particular 
to estimate the bounds on the parameter $q$ that depends on the  defects in the network,
we use a QIT concept called \textit{quantum telecloning} \cite{telecl}, which combines the phenomena of quantum teleportation \cite{tele} and quantum cloning \cite{clone}. 
While teleportation provides the means and fidelity with which a quantum state can be transferred to $\mathcal{M}$ parties using shared
pairwise
entanglement and classical communication, quantum cloning provides the optimal fidelity with which $\mathcal{M}$ copies 
of a quantum state can be prepared. Consider a site $a$ ($\in \mathcal{A}$) with $\mathcal{M}$ sites surrounding it, 
given by $\{b_i\}$  ($\in \mathcal{B}$), giving rise to $\mathcal{M}$ reduced states, $\rho_{ab_i}$. We suppose that an ancillary system in an 
arbitrary quantum state, $|\alpha\rangle$, is brought near the site $a$. This can be teleported to the site $b_i$, using the channel $\rho_{ab_i}$, 
with some optimal fidelity, $\mathcal{F}_{tele}$. If all $\rho_{ab_i}$'s are the same ($a$ has $\mathcal{M}$ equivalent neighbors $b_i$ in an 
isotropic lattice), then the state $|\alpha\rangle$ can be teleported to $\mathcal{M}$ sites with fidelity $\mathcal{F}_{tele}$. However, 
using an optimal cloning machine, $\mathcal{M}$ copies of a $d$-dimensional quantum state, $|\alpha\rangle$, can only be produced with a 
fidelity, 
\begin{equation}
\mathcal{F}_{clo} = \frac{2\mathcal{M}+(d-1)}{\mathcal{M}(d+1)}.
\end{equation}
Therefore, $\mathcal{F}_{tele} \leq \mathcal{F}_{clo}$, and we obtain an 
upper bound on the fidelity with which $\mathcal{M}$ copies of a quantum state can be remotely prepared. 

\begin{figure}[h]
 \includegraphics[angle=0,width=6cm]{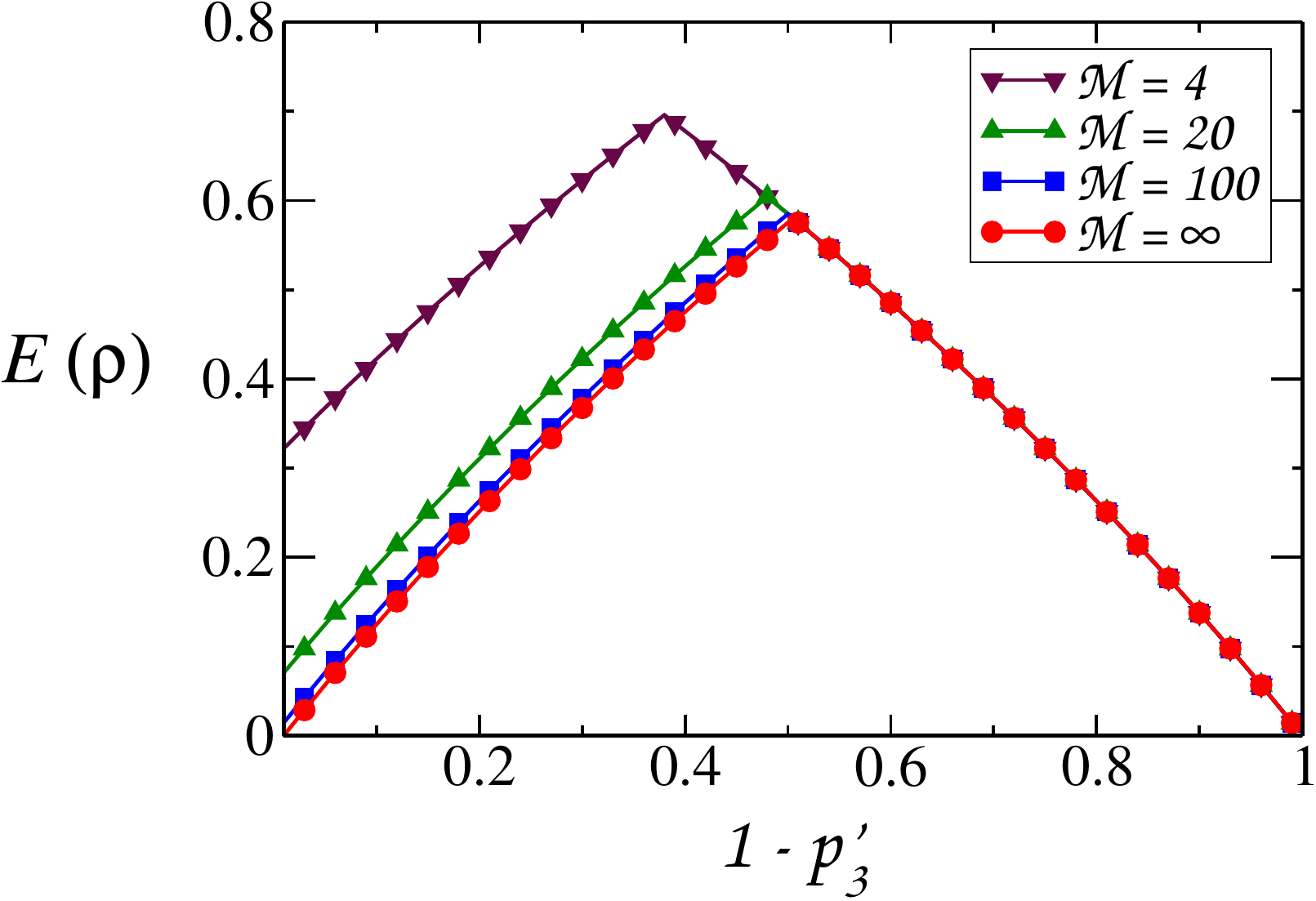} 
\caption{(Color online). Maximum permissible bipartite entanglement, $E(\rho)$. The plot shows the upper bound on BE due to quantum telecloning, quantified using LN, for 
$\rho^{(2)}_{\mathcal{A}\mathcal{B}_i}$, as a function of $1-p_3^\prime$. 
$E(\rho)$ decreases as $\mathcal{M}$ increases. Interestingly, in the absence of defects ($p_3^\prime$ = 1) no BE is present as $\mathcal{M} \rightarrow \infty$. The vertical axis is in ebits, while the horizontal one is dimensionless.}
\label{fig2}
\end{figure}

Suppose now that the quantum state of the system under consideration is the required resource for the remote protocol,
an unknown qutrit, $|\alpha\rangle$, is brought near site $a$, and 
$\mathcal{M}$ copies of it
are to be prepared at $\mathcal{M}$ sites, $\{b_i\}$, on the isotropic network. 
The $\mathcal{M}$ equivalent two-site density matrices, $\rho^{(2)}_{a b_i}$, are of the form given in Eq.~(\ref{two}). 
The fidelity of teleporting $|\alpha\rangle$ from near site $a$ to site $b_i$, is obtained from the maximal singlet fraction, $F$ = $\max{\langle \psi^s| \Lambda(\rho^{(2)}_{ab_i})|\psi^s\rangle}$, where the maximization is over all local operations and classical communication protocols, $\Lambda$, and where $|\psi^s\rangle$ is a maximally entangled state in $\mathbb{C}^d \otimes \mathbb{C}^d$ . The teleportation fidelity \cite{clone} is then given by 
\begin{equation}
\mathcal{F}_{tele} = {(Fd + 1)}/{(d+1)}.
\end{equation}
For qutrit, $d$ = 3, and $|\psi^s\rangle$ = $\frac{1}{\sqrt{3}}(|\zeta^\prime_1\zeta^\prime_2\rangle-|\zeta^\prime_1\zeta^\prime_2\rangle+|\zeta_0\zeta_0\rangle)$ and for $\Lambda$ as the identity operation,  $F^\prime$ = ${\langle \psi^s| \rho^{(2)}_{a b_i}|\psi^s\rangle} \leq F$. Using Eq.~(\ref{two}), for $\rho^{(2)}_{a b_i}$, we obtain 
\begin{equation}
F^\prime = {p^\prime_1}/{3} + ({p^\prime_3}/{3})[(3q+1)/{2}].
\end{equation}
Therefore, for $d$ = 3, we obtain
\begin{eqnarray}
\mathcal{F}_{tele} &\geq& \frac{(F^\prime d + 1)}{(d+1)} = \frac{p^\prime_1 +p^\prime_3[(3q+1)/2]+1}{4},
\label{ftele}\\
\mathcal{F}_{clo} &=& \frac{2\mathcal{M}+(d-1)}{\mathcal{M}(d+1)} = \frac{1}{2} + \frac{1}{2\mathcal{M}}. 
\label{fclone}
\end{eqnarray}
As discussed earlier, $\mathcal{F}_{tele} \leq \mathcal{F}_{clo}$, and using (\ref{ftele}) and (\ref{fclone}), we obtain an upper bound on the parameter $q$ in $\rho^{(2)}_{a b_i}$, given by
\begin{equation}
q \leq \frac{1}{3}\left(\frac{2}{p^\prime_3}-1\right) - \frac{2}{3p^\prime_3}\left(p^\prime_1-\frac{2}{\mathcal{M}}\right).
\end{equation}
As we know that the two-site density matrix is entangled if $q > 1/3$, 
the above relation provides us important indicators about the BE between two sites of the lattice. 
For relatively small number of defects in the lattice, $p^\prime_1 \ll p^\prime_3$, such that $p^\prime_1/p^\prime_3 \approx 0$. Hence, $q \leq (1/3)(1+\frac{4}{{p}'_3\mathcal{M}}+\delta)$, where $\delta$ = $2/p^\prime_3 - 2$, with $\delta \rightarrow 0$ as $p^\prime_3 \rightarrow 1$. Hence, the upper bound on $q$ decreases as the number of copies, $\mathcal{M}$, increases. 
For example, let us consider an isotropic 2D square lattice with a low number of defects. We consider the telecloning of a qutrit from a site to its four NNs ($\mathcal{M}$ = 4), so that the bound on $q$ for NN two-party reduced density matrices is given by $q \leq 1/3(1+\frac{1}{{p}'_3}) +\delta/3$. Hence, the NN state with highest BE for the isotropic 2D square lattice is given by Eq.~(\ref{two}), with $q$ = $1/3(1+\frac{1}{{p}'_3}) +\delta/3$. Similarly, let us consider the telecloning of a qutrit to $\mathcal{R}$ nodes ($\{b_1, b_2, \ldots, b_{\mathcal{R}}\}$), which are $x$ edges away from node $a$, such that $\mathcal{M}$ = $\mathcal{R}$. Alternatively, one may consider $\mathcal{R}$ to be the number of nodes contained in the area formed by concentric circles of radius $r > x$ and $r \leq x+x^\prime$, where $x^\prime \ll x$. For large $x$, all the $\mathcal{R}$ nodes can be considered as equidistant from node $a$. 
$\mathcal{R}$ increases with $x$, and we have $q \leq 1/3 +\delta/3$. We observe that as $\delta \rightarrow 0$, $q \leq 1/3$ and the states $\rho_{a b_i}^{(2)}$ (\(i=1,2,\ldots, \mathcal{R}\)) are separable, since $\rho_{a b_i}^{(2)}$ is then a mixture of three unentangled states. Figure~\ref{fig2} shows that the permissible upper limit on BE, as quantified by logarithmic negativity (LN) \cite{ppt,LN}, is finite for any two sites in the isotropic lattice in the presence of defects but decreases as 
$\mathcal{M}$ increases.

\section{Multipartite entanglement properties}
\label{multiartite entanglement}
\begin{figure}[t]
 \includegraphics[angle=0,width=4cm]{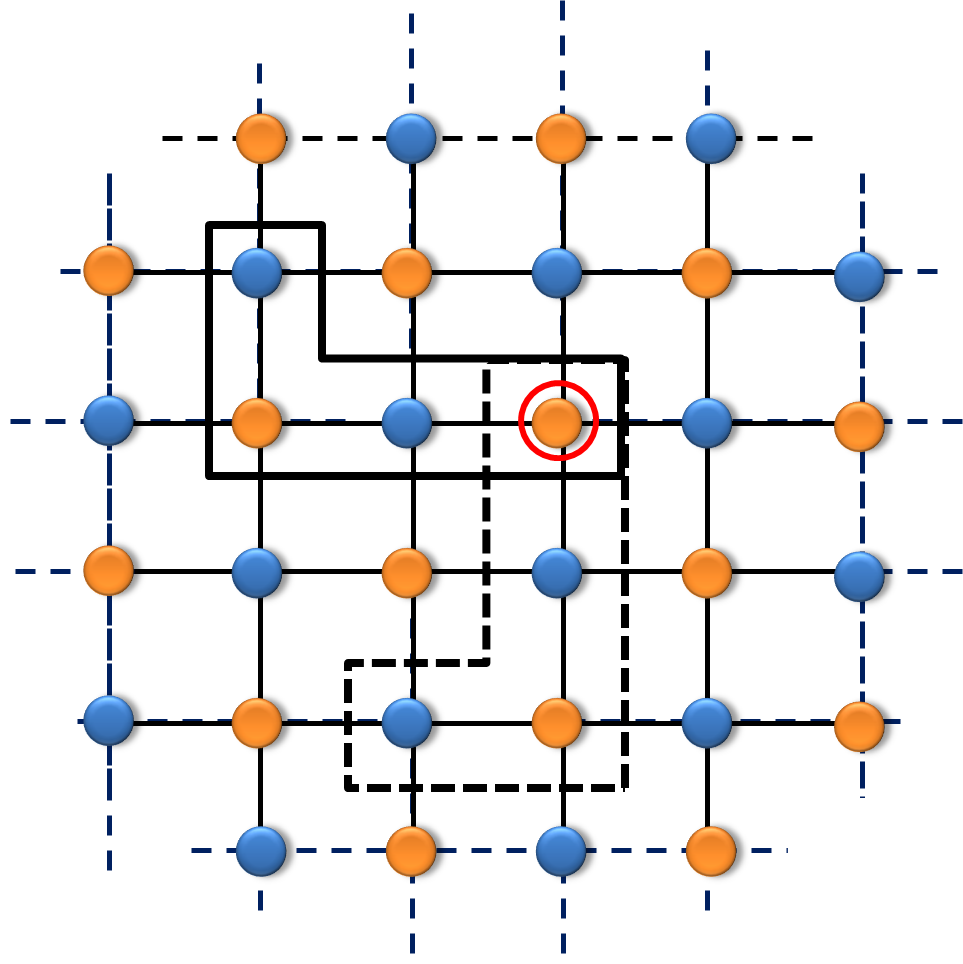} 
\caption{(Color online). Genuine multisite entanglement in isotropic lattice. 
Each node either contains a spin, which is a part of a dimer, or a defect. 
The spatially different but equivalent sets of $x$ (black-solid-line) and $x^\prime$ (black-dashed-line) nodes contain a common $y$ node (red circle). }
\label{fig3}
\end{figure}

We now 
show that an $\mathcal{N}$-spin network with $\mathcal{P}$ defective nodes, given by $|\bar{\Psi}\rangle_\mathcal{N}^\mathcal{P}$, such that $\mathcal{P} < \mathcal{N}$, is \textit{always} GM entangled. 
For a pure multiparty quantum state to be GM entangled, it must be entangled across all possible bipartitions of the system. This requires that the reduced density matrices across all possible bipartitions must necessarily be mixed. 
From Eqs.~(\ref{single}) and (\ref{two}), for $\mathcal{D} < 1$, we observe that the reduced single- and two-node reduced states are always mixed. Hence, the state $|\bar{\Psi}\rangle_\mathcal{N}^\mathcal{P}$ is entangled across all \textit{single:rest} and \textit{two:rest} bipartitions. Now we need to show that 
$|\bar{\Psi}\rangle_\mathcal{N}^\mathcal{P}$ is entangled across the other  possible bipartitions.
Consider the reduced state,  $\rho^{(x)} = \textrm{Tr}_{\bar{x}}[(|\bar{\Psi}\rangle\langle\bar{\Psi}|)_\mathcal{N}^\mathcal{P}]$, 
for an arbitrary but fixed set of $x$ nodes. Let us assume that $\rho^{(x)}$ is pure and thus $|\bar{\Psi}\rangle_\mathcal{N}^\mathcal{P}$ is 
separable along the $x:(\mathcal{N}-x)$ bipartition. Let $\{x\}$ = $\{\tilde{x}\} \cup y$, where $y$ is one specific node (among the 
\(x\) nodes), 
\(\{\tilde{x}\} = \{x\} \setminus y\), and \(\{\cdot\}\) denotes the set of the corresponding  lattice points. 
Since, 
the spin network is isotropic, there shall always exist an equivalent but spatially different set of $x^\prime$ nodes, such that both $x^\prime$ and $x$ 
contain an equal number of nodes and the node $y$ is common to both  sets. For an example, see Fig.~\ref{fig3}. 
Let us, similarly as before, consider $\{x^\prime\}$ = $\{\tilde{x}^\prime\} \cup y$, 
with \(\{\tilde{x}^\prime\} = \{x^\prime\} \setminus y\).
Now, 
$\rho^{(x^\prime)}$ is also  pure, 
by the symmetry of the  isotropic lattice. Applying 
the strong-subadditivity of $\mathcal{S}_\textrm{VN}$ ($S(\cdot)$) \cite{von}, we obtain
\begin{equation}
S(\rho^{(\tilde{x})})+S(\rho^{(\tilde{x}^\prime )}) \leq  S(\rho^{{\{\tilde{x}\}} \cup y})+ S(\rho^{{\{\tilde{x}}^\prime\}  \cup y}).
\label{VN}
\end{equation}
As $\rho^{(x)}$ and $\rho^{(x^\prime)}$ are pure, 
\(S(\rho^{{\{\tilde{x}\}} \cup y}) = S(\rho^{{\{\tilde{x}}^\prime\}  \cup y})=0 \).
Since $S(\cdot)$ is non-negative, $S(\rho^{({\tilde{x}})})$ = $S(\rho^{({\tilde{x}}^\prime)})$ = 0, 
and this implies that $S(\rho^{y})$ = 0, so that $\rho^{y}$ is pure. 
However, $y$ is a single node, and from Eq.~(\ref{single}), all single node reduced states are mixed, for $\mathcal{D} < 1$. 
This is a contradiction, proving that 
$\rho^{(x)}$ must be mixed for all sets of \(x\) nodes for any \(x>0\), and thus 
$|\bar{\Psi}\rangle_{\mathcal{N}-\mathcal{P}}$ is entangled across all $x:(\mathcal{N}-x)$ bipartitions. This proves that 
$|\bar{\Psi}\rangle_{\mathcal{N}-\mathcal{P}}$ is GM entangled for spin networks with finite defects. 
The above method can also be extended to prove that infinite spin networks are entangled across bipartitions by infinite lines. 
For example, the isotropic spin state, $|\bar{\Psi}\rangle$, defined on an infinite 2D lattice is always entangled across infinite lines on the lattice.
This proves that the entire multiparty state of the isotropic spin network is GM entangled, except in the extreme case where all 
spins are lost ($\mathcal{D}$ = 1), so that the state becomes a product of vacuum nodes.

The single- and two-node reduced states in Eqs.~(\ref{single}) and (\ref{two}), 
respectively, are defined for the qutrit quantum network with finite defects. Without defects, 
the network can be mapped to a tensor product of qubit spaces, spanned by $\{|\zeta^\prime_1\rangle,|\zeta^\prime_2\rangle\}$. The state $\rho^{(1)}_{a}$ is 
then  $I_2/2$ ($p_1$ = 0, $p_2$ = 1 in Eq.~(\ref{single})) and $\rho^{(2)}_{ab}$ is given by the Werner state, 
$\mathcal{W}(q)$ ($p^\prime_1$ = $p^\prime_2$ = 0, $p^\prime_3$ = 1). Since, $\rho^{(1)}_{a}$ is always mixed, 
using the approach discussed earlier, but applied to the case of zero defects, it can be shown that the spin network is always GM entangled. Finite defects in the network do not destroy the multiparty entanglement, but {for} the extreme case, where all spins are lost ($\mathcal{D}$ = 1).

In the absence of defects, the condition of BE between any two arbitrary nodes reduces to $q \leq (1/3)(1+2/\mathcal{M})$, since $\delta$ = 0. Again, considering the example of an isotropic 2D square lattice, in the limit $\mathcal{M} \rightarrow \infty$, we obtain $q \leq 1/3$, and the upper bound ensures that the system has no long-range BE \cite{aditi-prl}. However, in the presence of finite defects, the upper bound may allow a small but finite entanglement, as $q \leq (1/3)(1+ \delta)$.
Hence, presence of defects in quantum spin networks may not qualitatively affect the presence of GME in the system but, counterintuitively, may permit the presence of finite BE between 
moderately-distant sites, in contrast to the spin network with no defects.

\section{conclusions}
\label{discussion}In summary, our work aims at highlighting the response to defects of the distribution of entanglement,
both bipartite as well as multipartite, in a particular kind of quantum spin networks. Our results show that presence of finite defects do not affect the presence of GME in the network, while, interestingly, finite moderate-range BE may emerge. This shows that dimerized isotropic spin networks
provide a robust model for implementation of quantum protocols in presence of defects. 
Potential candidates for implementing such spin networks are spin-1/2 mixtures of ultracold atoms in optical lattices \cite{opt-lattice}. By tuning the tunneling between NN sites and the on-site interaction, at low temperatures, a favorable 
antiferromagnetic order and dimerization is achievable \cite{ol}. The defects in the network occur through controlled doping.
Alternatively, photonic lattices have also allowed for simulation of similar many-body networks \cite{photo1,photo2,photo3}.
Importantly, the spin network considered in our work is closely related to specific dimer models relevant to \textit{fault-tolerant computation} \cite{kit2003,nayak2008,wilz,sondhi, topo, res-topo} and to interpolations of projected entangled pair states~\cite{cirac0, cirac1, cirac2}, which are key resources for \textit{measurement-based quantum computation} \cite{mbqc,resource,review}.

\acknowledgements
The research of SSR was supported in part by the INFOSYS scholarship for senior students. HSD acknowledges funding by the Austrian Science Fund (FWF), 
project no. M 2022-N27, under the Lise Meitner programme. DR acknowledges support from the EU Horizon 2020-FET QUIC 641122.

 \end{document}